\documentstyle[openbib,12pt]{article}
\begin{document}
\setcounter{page}{1}
\title{\mbox{ } \\[-3cm]
{\footnotesize\hspace*{\fill}July - 1993\\
[-1mm] \footnotesize\hspace*{\fill}DFFCUL 01-07/1993  \\
 [-4mm] \footnotesize\hspace*{\fill} DF/IST 12.93}
\\[1cm]
Decoherence of Homogeneous and Isotropic Geometries
in the Presence of Massive Vector
Fields\thanks{Talk presented at the $3^{{\rm rd}}$ National
Meeting of Astronomy and Astrophysics, July 1993, IST,
Lisboa, Portugal.}}
\author{{\large\bf  O. Bertolami}
\thanks{Address after August 1993: CERN, Theory Division,
CH-1211 Gen\`eve 23, Switzerland.} \\
Departamento de F\'{\i}sica,
Instituto Superior T\'ecnico \\
Av. Rovisco Pais 1, 1096 Lisboa Codex, PORTUGAL
\and
{\large\bf
\underline{P.V. Moniz}}\thanks{Address after October 1993: University of
Cambridge, DAMTP,
Silver Street, Cambridge, CB3 9EW, UK .}
\thanks{E--MAIL: prlvm10@amtp.cam.ac.uk}
\thanks{Work supported in
part by the J.N.I.C.T.
graduate scholarship BD/138/90-RM and by funds provided by
the project STRIDE/FEDER JN.91.02
(D.F.F.C.U.L. -- J.N.I.C.T./C.E.R.N.)}
\\ Departamento de F\'{\i}sica, Universidade de
Lisboa \\ Campo Grande, Ed. C1, piso 4, 1700-Lisboa,
PORTUGAL}
\date{{\sf PACS numbers: 04.60.+n; 98.80.-Cq,-Dr}}
\maketitle

\vspace{-0.8cm}

\begin{abstract}
Decoherence of Friedmann-Robertson-Walker (FRW)
geometries due to
massive vector fields with $SO(3)$ global symmetry is
 discussed in the context of Quantum Cosmology.
\end{abstract}

\indent

The Universe on the large scale behaves classically to a
high degree of accuracy.  On the other
hand,  all matter in the Universe is
ultimately described
by quantum fields, and one expects
 the gravitational field will
eventually be described in a similar way.
Progress in this direction has lead in recent decades
to an extensive research \cite{hh,hallije}
about the quantization of the {\em whole} Universe,
assuming it can be described by a single wave function
$\mbox{\boldmath $\Psi$}$.
A  crucial issue
 in quantum cosmology
is  to understand how the classical
behaviour can be recovered in order to
make predictions \cite{halli3}-\cite{hg}.

The conditions that must be
satisfied in order that a system may be regarded as
classical are the following \cite{halli2}:
Firstly, its  the evolution
should be driven by classical laws, which means that
from the wave function  $\mbox{\boldmath $\Psi$}$
a strong {\em correlation}
between the configuration space coordinates and the
conjugate momenta
should exist. The second condition involves the notion of
{\em decoherence}, meaning that any  interference between
different quantum states should vanish.

The  decoherence process may emerge in the following way
\cite{kif2}-\cite{zu1}. One considers that
the original system is part of a more complex world
and interacts with another subsystem, usually
designated
as  {\em environment} and which is formed by
 unobserved
(or irrelevant) degrees of freedom. Then, under
some circunstances,
quantum interference effects on the original system can be
supressed by  the interaction with the environment.
Recent works \cite{hu4,hu2}  shows that the
two above mentioned criteria for the retrieval of classical behaviour,
i.e.,
correlation and decoherence, must be considered as
{\em inter}dependent aspects of the quantum to
classical transition and  should be analysed altogether.
To be more precise,
if one adopts the conditional probability
interpretation, according to which only
 peaks on the wave function, or on a probability
distribution function derived from $\mbox{\boldmath
$\Psi$}$, are tantamount to predictions in a
quantum cosmological scenario \cite{hh,halli3},
then a more careful   analysis indicates
\cite{hu4} that
the  wave function or
probability  distribution
does not a have a  peak around a
single
trajectory and hence, does not exhibit the
expected classical correlations
unless
some kind of coarse-graining is  introduced.
Thus, a system  coupled to an
environment induces it to  decohere.

Most of the literature
 about the
 retrieval of classical behaviour in quantum
cosmology  considers
minisuperspace models  in the presence of
scalar fields \cite{halli3}-\cite{hg}
and as environment   inhomogenous modes either of
 the gravitational or of the
matter fields\cite{kif2,halli1}.

In this work we propose an alternative
model which possesses  a non-Abelian global
symmetry,
and where the scalar field is replaced by
a massive vector field. The main reason to consider this model
  is to achieve an increasingly
realistic description of the early Universe incorporating
particle-physics-motivated
quantum fields, such as vector fields which acquire their
masses through spontaneous symmetry breaking mechanisms
 (cf.ref.\cite{cambada}).
Furthermore,  the presence of a mass term
is of crucial importance in the quantum cosmological context
since for  the massless case the
conformal invariance of the Yang-Mills action
gives origin
 to a Wheeler-DeWitt equation that allows for a
 decoupling  of
the gravitational and gauge degrees of freedom
\cite{moob}
(similarly to the case of a free massless conformally invariant scalar field
\cite{hh}). Hence,
 an
interaction process could not in that situation induce the
minisuperspace variables to decohere. The presence of a mass term breaks the
conformal invariance,
thus allowing interaction terms between
the gravitational and gauge degrees of freedom
and providing a scale to which the decoherence process
can  be studied.

The action of a model with a non-Abelian global
symmetry with a
massive vector field is  given by
\begin{equation}
{\cal S} =
\int_{{\cal M}^4}
d^4 x \: \sqrt{-\det \mbox{\boldmath $g$}}\:
\left\{\frac{M_{{\rm Pl}}^2}{16\pi}
{\cal R} + \frac{1}{8\varepsilon^2}
{\rm Tr}(\mbox{\boldmath $F$}_{\mu\nu}
\mbox{\boldmath $F$}^{\mu\nu}) +
\frac{1}{2} m^2
{\rm Tr}(\mbox{\boldmath $A$}_{\mu}\mbox{\boldmath
$A$}^{\mu})
\right\} -
\frac{M_{{\rm Pl}}^2}{8\pi}\int_{\partial{\cal M}^4}
d^3 x \: \sqrt{-\det \mbox{\boldmath $h$}}\:\: {\cal K},
\label{eq:act1}
\end{equation}
where $\mbox{\boldmath $g$}$ is the Lorentzian metric, with
signature $(-,+,+,+)$ in the four dimensional manifold
${\cal M}^4$,
$\mbox{\boldmath $h$}$ is
the induced spatial metric on the spatially hypersurfaces,
   ${\cal R}$ and ${\cal K}$ are the scalar curvature
corresponding to
$\mbox{\boldmath $g$}$ and the trace of the extrinsic
curvature ${\cal K}^{\mu}_{\nu}$ of ${\cal M}^4$,
respectively.
The constant $\varepsilon$ denotes a coupling constant,
$M_{{\rm Pl}}$  the Planck mass,
$m$
 the mass of the vector field $\mbox{\boldmath $A$}_{\mu}$
and $\mbox{\boldmath $F$}_{\mu\nu}$ is the usual field
strenght tensor.

We are interested  in spatially closed
FRW cosmologies associated with the action
(\ref{eq:act1}) for the case
where the  internal global symmetry is
$\hat{G} = SO(3)$.
The most general form of a
metric that is spatially homogenous and isotropic in
a ${\cal M}^4 = I\!\!R \times S^3$ topology is given by
the FRW ansatz
\begin{equation}
\mbox{\boldmath $g$} =
\sigma^2(-N^2(t) dt^2 + a^2(t) \Sigma_{a=1}^{3}
\mbox{\boldmath $\omega$}^a \otimes
\mbox{\boldmath $\omega$}^a),
\label{eq:met}
\end{equation}
where $N(t)$ and $a(t)$ are respectively
the lapse function and the
scale factor,
$\sigma^2 = \frac{2}{3\pi M_{{\rm Pl}}^2}$ and
$\mbox{\boldmath $\omega$}^a$ are left-invariant
1-forms in $SU(2) \stackrel{{\rm diff}}{\simeq}
S^3$ satisfying
$d\mbox{\boldmath $\omega$}^a =
-\epsilon_{abc} \mbox{\boldmath $\omega$}^b
\wedge \mbox{\boldmath $\omega$}^c$.

For the vector field we make the ansatz
\begin{equation}
\mbox{\boldmath $A$} =
\frac{1}{2}\left[1 + \sqrt{\frac{2\alpha}{3\pi}}
\tilde{\chi}_{0}(t)\right]{\cal T}_{bc}\epsilon_{bac}
\mbox{\boldmath $\omega$}^{a},
\label{eq:ansatz2}
\end{equation}
where ${\cal T}_{bc}$ are the generators of $\hat{G}$,
$\tilde{\chi}_0$ is  an arbitrary function of time
and $\alpha \equiv \frac{\varepsilon^2}{4\pi}$.
The ansatz (\ref{eq:ansatz2}) follows the method presented
in ref.\cite{cambada},
and corresponds to a vector field which is
homogeneous and isotropic (i.e., $SO(4)$-invariant,
where $G=SO(4)$ is the group of isometries of $S^3$)
up to a compensating internal global transformation.
This means that the energy-momentum tensor obtained from the
action (\ref{eq:act1}) and using the ansatz
(\ref{eq:ansatz2}) is of the form of a perfect fluid.

When the transition from quantum to classical states is
studied in a system with
scalar fields, one usually considers a perturbed Friedmann model and proceeds
to an expansion of the  scalar fields
in terms of spherical harmonics
\cite{hallije},\cite{kif2}-\cite{hg}. In our model, however,
we have a vector field with only spatial components and
therefore it should be expanded in terms of spin-1
spinor hyperspherical harmonics (cf. ref. \cite{dowk} and
references therein).
In order to get some insight we shall instead discuss here
an approach that consists in taking the vector field in the
following form
\begin{equation}
\mbox{\boldmath $A$} =
\frac{1}{2}\left[1 + \sqrt{\frac{2\alpha}{3\pi}}
\left\{\chi_{0}(t) +
\Sigma_{n,l,m}f_{nlm}(t)Q^{n}_{lm}
(\chi,\theta,\varphi)\right\}
\right]{\cal T}_{bc}\epsilon_{bac}
\mbox{\boldmath $\omega$}^{a},
\label{eq:ansatz2a}
\end{equation}
where $Q^{n}_{lm}
(\chi,\theta,\varphi)$ are spherical
harmonics on $S^3$ and
 $n=2,3,\ldots,\infty$; $l=0,1,\ldots,n-1$;
$m=-l,\ldots,l$.
The scale factor $a(t)$, the function $\chi_0(t)$
 and the infinite modes $f_{nlm}$ form the ``superspace".
Reduction to a minisuperspace is performed
 truncating the
higher modes ($n\geq 2$) which will play
the role of  the environment.
The assumption we are making here is that this
simplification contains the main features of the full model.
A more complete account of this problem will be discussed
elsewhere \cite{obpvm93}.

Substituting (\ref{eq:met}) and
(\ref{eq:ansatz2a}) into (\ref{eq:act1})
and remembering that the integration over
spatial coordinates gives the volume $2\pi^2$
for $S^3$ ($a=1$),
one obtains an
effective action:
\begin{eqnarray}
{\cal S}^{{\rm eff}} & = &
\frac{1}{2}\int \:d \eta\left\{
-\left(\frac{d a}{d \eta}\right)^2  + a^2
+ \left(\frac{d \chi_0}{d \eta}\right)^2
- \frac{2\alpha}{3\pi }\left[\chi_0^2 -
\frac{3\pi}{2\alpha}\right]^2 - \right.
\nonumber \\
& &
\frac{16\mu^2 a^2}{3}\left(\sqrt{\frac{3\pi}{2\alpha}}
+ \chi_0^2\right)^2 + \frac{1}{2\pi^2} \Sigma_{n}
\left(\frac{d f_n}{d \eta}\right)^2 -
\frac{2}{3\pi^2 }\Sigma_{n}f_{n}^2\left(n^2 -
\frac{5}{3}\right) - \frac{2\mu^2 a^2}{\pi}
\Sigma_{n}f_{n}^2 -
\nonumber \\
& &
\left. \frac{2\alpha}{\pi^3 }\chi_0^2
\Sigma_{n}f_{n}^2 - \frac{4\alpha}{3\pi^2 }
\chi_0 a_{nmp} f_n f_m f_p
- \frac{\alpha}{3\pi^3 }
b_{nmpq} f_{n} f_{m} f_{p} f_{q}
\right\},
\label{eq:act2a}
\end{eqnarray}
where the conformal time $d\eta = \frac{dt}{a(t)}$
has been used and
\begin{equation}
a_{nmp}  \equiv
\int_{S^3} d \Omega Q^n Q^m Q^p, \:\:
b_{nmpq}  \equiv
\int_{S^3} d\Omega Q^n Q^m Q^p Q^q,
\label{eq:help5b}
\end{equation}
where $d\Omega = \sqrt{\det \mbox{\boldmath $\Omega$}}
d\chi d\theta d\varphi$, $\mbox{\boldmath $\Omega$}$
being  the metric on $S^3$,
$\mu^2 \equiv \frac{m^2}{M_{{\rm Pl}}^2}$,
and the   subscripts on the $f$'s and $Q$'s
collectivelly denote the quantum numbers $(nlm)$.

Let us now turn to the derivation of the Wheeler-DeWitt
equation. The canonical momenta conjugate to
$a(t)$, $\chi_0$ and $f_n$ are given by
\begin{equation}
\pi_a = \frac{\partial {\cal L}^{{\rm eff}}}
{\partial (d a/d \eta)} =
- \frac{ d a}{ d \eta},\:
\pi_{\chi_0} = \frac{\partial {\cal L}^{{\rm eff}}}
{\partial (d \chi_0/d \eta)} =
 \frac{ d \chi_0}{ d \eta},\:
\pi_{f_n} = \frac{\partial {\cal L}^{{\rm eff}}}
{\partial (d f_n/d \eta)} =
\frac{1}{2\pi^2} \frac{ d f_n}{ d \eta}.
\label{eq:help6}
\end{equation}
The Hamiltonian constraint, ${\cal H}=0$,
 can be written as
\newpage

\begin{eqnarray}
& . & \frac{1}{2}\left\{-\pi_a^2  - a^2
+ \pi_{\chi_0}^2
+ \frac{2\alpha}{3\pi }\left[\chi_0^2 -
\frac{3\pi}{2\alpha}\right]^2 -
\frac{16\pi\mu^2 a^2}{3\pi}\left(\sqrt{\frac{3\pi}{2\alpha}}
+ \chi_0^2\right)^2 + \right.
\nonumber \\
& &
2\pi^2 \Sigma_{n}\pi_{f_n}^2 +
\frac{2}{3\pi^2 }\Sigma_{n}f_{n}^2\left(n^2 -
\frac{5}{3}\right) + \frac{2\mu^2 a^2}{\pi}
\Sigma_{n}f_{n}^2 +
\nonumber \\
& &
\left. \frac{2\alpha}{\pi^3 }\chi_0^2
\Sigma_{n}f_{n}^2 + \frac{4\alpha}{3\pi^2 }
\chi_0 a_{nmp} f_n f_m f_p
+ \frac{\alpha}{3\pi^3 }
b_{nmpq} f_{n} f_{m} f_{p} f_{q} \right\}
= 0 .
\label{eq:hamilt}
\end{eqnarray}
Quantization proceeds by promoting the
canonical conjugate momenta in
(\ref{eq:help6}) into operators, in the following way
\begin{equation}
\pi_a = -i\frac{\partial}{\partial a},\:
\pi_{\chi_0} = -i\frac{\partial}{\partial \chi_0},\:
\pi_{f_n} = -i\frac{\partial}{\partial f_n},\:
\pi_a^2 = -a^{-p}\frac{\partial}{\partial a}
\left(a^p \frac{\partial}{\partial a}\right),
\label{eq:help7}
\end{equation}
where the last substitution
 reflects the operator ordering ambiguity
being $p$  a real constant.
The Wheeler-DeWitt equation is then found to be
\begin{equation}
\frac{1}{2}\left[a^{-p}\frac{\partial}{\partial a}
\left(a^p \frac{\partial}{\partial a}\right)
- \frac{\partial^2}{\partial \chi_0^2}
-  2\pi^2 \Sigma_{n}
\frac{\partial^2}{\partial f_n^2}
+ {\cal V}_0(a, \chi_0) +
{\cal V}(a,\chi,f_n)
\right]
\mbox{\boldmath $\Psi$}[a, \chi_0,
f_n]
= 0.
\label{eq:wdw}
\end{equation}
where
\begin{eqnarray}
{\cal V}_0 & = & -a^2
+ \frac{2\alpha}{3\pi }\left[\chi_0^2 -
\frac{3\pi}{2\alpha}\right]^2 +
\frac{16\pi\mu^2 a^2}{3\pi}\left(\sqrt{\frac{3\pi}{2\alpha}}
+ \chi_0^2\right)^2  \label{eq:pot1} \\
{\cal V} & = &
 \frac{2\alpha}{\pi^3 }\chi_0^2
\Sigma_{n}f_{n}^2
+
\frac{2}{3\pi^2 }\Sigma_{n}f_{n}^2\left(n^2 -
\frac{5}{3} + 3\pi\mu^2a^2\right)
+ \frac{4\alpha}{3\pi^2 }
\chi_0 a_{nmp} f_n f_m f_p + \frac{\alpha}{3\pi^3 }
b_{nmpq} f_{n} f_{m} f_{p} f_{q}
\label{eq:pot2}
\end{eqnarray}

It is interesting to compare
this result with the  case of a FRW minisuperspace model
with a massive scalar field with a self-interaction term
$\lambda\Phi^4$ and a conformal coupling term $\frac{1}{12}
{\cal R} \Phi^2$ \cite{thehu1}. The
Wheeler-DeWitt equation in this case is given by
\begin{equation}
\left[\frac{1}{2}a^{-p}\frac{\partial}{\partial a}
\left(a^p \frac{\partial}{\partial a}\right)
- \frac{1}{2}\frac{\partial^2}{\partial \Phi_0^2}
+ \frac{1}{2}\Sigma_{n}
\frac{\partial^2}{\partial f_n^2}
+  {\cal V}_0 + {\cal V}\right]
\mbox{\boldmath $\Psi$}[a, \Phi_0,
f_n]=0,
\label{eq:wdw2a}
\end{equation}
where
\begin{eqnarray}
{\cal V}_0 & = & -\frac{1}{2}a^2
+ \frac{1}{2}m^2a^2\Phi_0^2 +
\frac{\lambda}{4!}\Phi^4_0 \label{eq:pot1a} \\
{\cal V} & = & \frac{6\lambda}{4!}
\Sigma_k\Phi^2_0f_k^2 +
\frac{1}{2}\Sigma_k(k^2 + m^2a^2)f_k^2 +
\frac{\lambda}{4!} a_{klm}
f_kf_lf_l\Phi_0 +
\frac{\lambda}{4!}b_{klmn}
f_kf_lf_mf_n.
\label{eq:pot2a}
\end{eqnarray}

One clearly sees that  both models are fairly similar.
In particular, if one drops the
$a_{klm}$ and $b_{klmn}$ terms, retaining
in (\ref{eq:pot2}) and  (\ref{eq:pot2a})
only
the terms quadratic in the $f_n$'s,
one gets a
{\em non-linear}
interaction
between the lowest mode ($n=1)$ with the
the environment, i.e, the higher modes.
Phenomena such as dissipation, fluctuation,
decoherence, back-reaction and particle creation
are particularly interesting in this last minisuperspace
model, suggesting they may be also worth
studying in our model (cf.refs.\cite{hu3}).
It is  important to stress that up to
ref.\cite{thehu1} the generality of the
 models  that were considered
containing only  free scalar fields
(in some cases with a conformal coupling term)
on a FRW
background presented no
interaction between the lowest
and  the
higher modes.

However, there are
some important differences between our model and the one in
ref. \cite{thehu1}. In the first place, the
effective potential of self-interaction for $\chi_0$ in
(\ref{eq:pot1}) is of the  anharmonic  type \cite{moob}.
Moreover,
 the effective action (\ref{eq:act2a})
 describes the dynamics of the early Universe
via  variational principles in  a setting
where earlier
 than the inflationary period the Universe
is dominated by
gravity and non-Abelian vector fields, being therefore
more realistic. In this
context,
our approach represents a step forward towards the
 understanding of the transition from    quantum
to
 classical regime in the very early universe.

{\large\bf Acknowledgments}

The authors  gratefully acknowledge
B.L. Hu  for  enlightening  discussions and
suggestions, L. Garay and C. Kiefer
for   pleasant conversations and    J. Mour\~ao
for  valuable  comments.

\end{document}